\documentclass[reprint,aps,amsmath,amssymb,twoside,prd,showkeys,superscriptaddress]{revtex4-1}
\pdfoutput=1
\usepackage{verbatim}
\usepackage{comment}
\usepackage{graphicx}
\usepackage[T1]{fontenc}
\usepackage{epsfig}
\usepackage{bm}
\usepackage{amssymb}
\usepackage{float}
\usepackage{amsmath}
\usepackage{subfigure}
\usepackage{booktabs}
\usepackage{multirow}
\usepackage{array}
\usepackage{natbib}
\usepackage{dcolumn}
\usepackage{cancel}
\usepackage[colorlinks]{hyperref}
\usepackage[usenames,dvipsnames]{color}
\hypersetup{
    breaklinks=true,
    pdfstartview={FitH},
    colorlinks=true,
    linkcolor=blue,
    citecolor=red,
    filecolor=magenta,
    urlcolor=blue,
    anchorcolor=green,
    linktocpage=true
}
\usepackage{verbatim}
\usepackage{comment}
\usepackage{etoolbox}

\usepackage{longtable}

\def\doi{http://doi.org}

\newcommand{\HCd}{\mathcal{H}}

\def\HCdt0{\tilde{\HCd}_{0}}

\newcommand{\affLib}{Leibniz-Institut fur Astrophysik Potsdam (AIP), An der Sternwarte 16, 14482 Potsdam, Germany}
\newcommand{\affdel}{Dipartimento di Fisica e Astronomia, University Of Catania, Viale Andrea Doria 6, 95125, Catania, Italy}

\begin{document}

\title{Islands in Simulated Cosmos: Probing the Hubble Flow around Groups and Clusters}

\author{David Benisty}
\email{benidav@aip.de}
\affiliation{\affLib}

\author{Antonino Del Popolo}
\email{antonino.delpopolo@unict.it}
\affiliation{\affdel}

\begin{abstract}
 {The local Hubble flow provides a valuable probe of the transition between cosmic expansion and nonlinear gravitational dynamics. On large scales, galaxies follow the linear Hubble law, but within group- and cluster-sized environments, gravitational interactions generate substantial deviations. Using the IllustrisTNG cosmological simulations, we test whether dark energy leaves measurable signatures in the local velocity--radius relation. We model the kinematics with extensions of the Lema\^{\i}tre--Tolman framework and use Bayesian inference to recover halo masses and the Hubble constant $H_0$. The fits exhibit systematic biases: halo masses are recovered with a median ratio $\big\langle M_{\rm fit}/M_{\rm true}\big\rangle = 0.991 \pm 0.148$, while the inferred expansion rate peaks at $\big\langle H_{0,\text{fit}}/H_{0,\text{True}} \big\rangle =1.01 \pm 0.14$. Although both mass and $H_0$ can be constrained from the local flow, the different model variants, as the angular momentum, friction-like terms, or dark energy, remain statistically indistinguishable given the intrinsic environmental variance. Our results demonstrate both the potential and the fundamental limitations of using local kinematics as a precision diagnostic of dark energy.}
\end{abstract}


\maketitle

\section{Introduction}
\label{sec:intro}

 {The dynamics of galaxy groups and clusters are fundamentally defined by a cosmic tug-of-war: the inward pull of nonlinear gravitational collapse competing against the outward push of local cosmic expansion. During the formation of these structures, a natural kinematic segregation occurs. Matter within the inner, overdense regions breaks away from the background expansion to virialize into a gravitationally bound halo. In contrast, material at greater distances continues to expand alongside the local Hubble flow~\cite{bib:Sandage1986,Weinberg:1987dv,Peirani:2005ti,Penarrubia:2014oda}. The boundary between these two dynamical regimes imprints a distinct signature on the system's velocity--distance relation.}

 {This transitional zone is characterized by the turnaround radius, a critical metric that encodes direct information about both the enclosed dynamical mass and the broader cosmological model~\cite{bib:Sandage1986,Pavlidou:2004vq,Pavlidou:2013zha}, that reads:}
\begin{equation}
M = \frac{\pi^2 r_0^3}{8 G t_U^2},
\label{eq:R0_def}
\end{equation}
 {where $r_0$ is the turnaround radius, $M$ is the mass of the halo, and $t_U$ is the age of the Universe. Ref.~\cite{Peirani:2005ti} demonstrated that a system's total mass and the Hubble constant can be derived from the peculiar velocities of nearby galaxies relative to the Hubble flow. Since this foundational work, the theoretical framework has been significantly refined. By incorporating generalized dark matter halo profiles, anisotropic infall, and tidal interactions, subsequent studies have successfully reproduced the complex velocity dispersions observed in real galaxy groups~\cite{Peirani:2008qs,Teerikorpi:2010zz,Penarrubia:2015hqa,Baushev:2019mrg,Baushev:2020dnq,DelPopolo:2021hkz,DelPopolo:2022sev,Wagner:2025wrp,Benisty:2025tct}.}

 {Galaxy groups such as the Local Group~\cite{Karachentsev:2008st,Penarrubia:2014oda,Makarov:2025} and other nearby associations~\cite{karachentsev2002,Karachentsev:2006ww,Faucher:2025blj} typically exhibit turnaround radii of order $\sim 1$--$2\,\mathrm{Mpc}$ and total masses of a few $\times 10^{12}\,M_\odot$. In contrast, galaxy clusters probe significantly larger scales, with turnaround radii of $\sim 4$--$7\,\mathrm{Mpc}$ and masses ranging from $10^{14}$ to several $10^{15}\,M_\odot$~\cite{Kim:2020gai,Sorce:2016yok,Karachentsev:2010nw,Nasonova:2011md,Benisty:2025tct}. This wide dynamical range makes the turnaround scale a powerful probe of nonlinear gravitational collapse over more than three orders of magnitude in mass.}

 {The role of dark energy in shaping local velocity fields remains an open question. Ref.~\cite{Hoffman:2007qe} argued that dark energy cannot be robustly detected from local Hubble flow measurements alone, while Refs.~\cite{Peirani:2008qs,DelPopolo:2022sev} explored fits to alternative dark energy models using velocity--distance relations around nearby structures. A key open question is therefore whether the natural halo-to-halo variance in a realistic simulation is small enough to discriminate between these models.}

\begin{table*}[htbp]
\centering
\label{tab:model_comparison}
\begin{tabular}{l l c c c}
\hline
\textbf{Model} & \textbf{Key Assumptions \& Physical Effects} & \textbf{A} & \textbf{b} & \textbf{n} \\
\hline
MLT & No angular momentum ($J=0$) and no dynamical friction ($\eta=0$) & 1.013 & 1.4054 & $0.96$ \\
JLT & Includes specific angular momentum ($J$) but neglects dynamical friction ($\eta=0$) & 0.8016 & 1.3759 & $0.75$ \\
J$\eta$LT & Accounts for both angular momentum and dynamical friction ($\eta$) & 0.6639 & 1.3436 & $0.91$ \\
TA & Based on timing argument for dwarf galaxies; compatible formulation & 1.1 & 1.41 & 0.5 \\
\hline
\end{tabular}
\caption{ { {\it{Comparison of Lema\^{\i}tre Tolman model variants and their parameters. Different assumptions about dark energy and additional physical effects (angular momentum, dynamical friction) lead to systematically different values of the coefficients $A$, $b$, and $n$~\cite{DelPopolo:2022sev}. The transition from MLT to J$\eta$LT reflects increasing physical complexity. The TA model fixes $n = 0.5$ following the timing argument~\cite{Penarrubia:2014oda}.}}}}
\end{table*}

 {In this paper, we use the IllustrisTNG cosmological simulations~\cite{Nelson:2018uso} to address this question quantitatively. While previous applications of the Lema\^{\i}tre--Tolman (LT) framework to observational data provided constraints based on statistical uncertainties, testing the method in a controlled simulation environment allows us to robustly quantify its \textit{systematic} uncertainties. By using a high-resolution simulation as a cosmic laboratory, we evaluate the impact of oversimplifications in the analytical model, such as spherical symmetry and isolated dynamics, when confronted with realistic environmental and nonlinear effects. Crucially, our results demonstrate that the local Hubble flow method can robustly only constrain the halo mass $M$ and the Hubble constant $H_0$, and cannot constrain the dark energy equation of state due to  statistical errors of the velocity distance relation.}

This paper is organized as follows. In Section~\ref{sec:test_particle}, we derive the test-particle equations of motion in an expanding universe and obtain the velocity--radius relations that form the basis of our analysis. Section~\ref{sec:sim} describes the simulation data and the selection of isolated halos from IllustrisTNG. In Section~\ref{sec:results}, we present fits to the velocity - radius relation and quantify the recovery of halo masses and the Hubble constant, including systematic uncertainties. Finally, Section~\ref{sec:discussion} discusses the implications for dark-energy studies and the extent to which local velocity fields can complement traditional cosmological probes.

\section{Hubble Flow Parametrization}
\label{sec:test_particle}

The motion of a system embedded in an expanding cosmological background can be treated, in the weak field limit, as a test particle (with the reduced mass) moving in an effective potential that contains both the Newtonian central attraction and the background cosmological acceleration. A convenient spacetime describing a central mass embedded in an FLRW universe is the McVittie metric~\citep{McVittie:1933zz, Kaloper:2010ec}; in the weak--field, slow--motion limit this leads to a modified radial equation of motion of the form (see also \cite{Sereno:2007tt, Faraoni:2007es, Nandra:2011ug}):
\begin{equation}
\ddot{r} = -\frac{G M}{r^2} +  \frac{l^2}{r^3} + \frac{\ddot{a}}{a} - \eta\,  \dot{r},
\label{eq:eomDaDt_recap}
\end{equation}
where $r(t)$ is the physical separation of the two bodies, $M$ the total (enclosed) mass, $G$ Newton's constant, $l$ the angular momentum per unit reduced mass ($l \equiv r_0 v_{\rm tan}$), $a(t)$ the cosmological scale factor, and $\eta$ a phenomenological dynamical friction coefficient~\cite{DelPopolo:2022sev}. For a spatially flat $\Lambda$CDM cosmology the cosmic acceleration is:
\begin{equation}
\frac{\ddot{a}}{a} = H_0^2 \left[-\frac{1}{2}\Omega_{m,0} a^{-3} + \Omega_{\Lambda,0}\right],
\label{eq:ddota_over_a}
\end{equation}
where $H_0$ is the present-day Hubble constant and $\Omega_{m,0},\Omega_{\Lambda,0}$ are the matter and dark--energy density parameters. To compare with observations, it is convenient to define the peculiar velocity,
\begin{equation}
v_{\rm pec}(t) \equiv v_{\rm rad}(t) - H(t)\,r(t),
\label{eq:vpec_def}
\end{equation}
where $H(t)=\dot a/a$ is the instantaneous Hubble parameter.  {Integrating Eq.~(\ref{eq:eomDaDt_recap}) from $v_{\text{pec}} = 0$, and using the turnaround scale $r_0$ to non-dimensionalize, yields an empirical velocity-radius law~\cite{Peirani:2005ti}:}
\begin{equation}
v(r) = -A \frac{H_0}{r^n}
       \left(\frac{GM}{H_0^2}\right)^{\frac{n+1}{3}}
       + b H_0 r,
\label{eq:vR_general}
\end{equation}
where $b$, $A$, and $n$ are dimensionless constants.  {The first term captures nonlinear infall toward the halo, while the second term captures the background Hubble expansion; the interplay between the two defines the turnaround radius via $v(r_{\rm ta}) = 0$. Different physical assumptions, about the dark energy model, the angular momentum, or the friction term, lead to specific numerical values of $A$, $b$, and $n$, as summarized in Table~\ref{tab:model_comparison} and derived explicitly in Ref.~\cite{DelPopolo:2022sev}.}

 {The dark energy contribution $\Omega_{\Lambda,0}$ explicitly enters the equation of motion through the cosmic acceleration term $\ddot{a}/a$: different dark energy models (or models with a dynamical equation of state $w \neq -1$) modify this term, which in turn changes the shape of the velocity, radius relation and, consequently, the fitted coefficients $A$, $b$, and $n$.}

 {The transition from MLT to J$\eta$LT shows a systematic decrease in $A$ and $b$, reflecting the increasing physical complexity of the models. The full J$\eta$LT model further includes dynamical friction, whose dissipative effect lowers the coefficients and slows the collapse. The approximate Timing Argument is another special case of Eq.~(\ref{eq:vR_general}) with fixed $n=0.5$~\cite{Penarrubia:2014oda}:}
\begin{equation}
v(r) = \left(1.2 + 0.31\Omega_\Lambda \right) H_0 r - 1.1 \sqrt{\frac{G M}{r}}.
\end{equation}
 {In our analysis we infer the parameters from the TNG simulation to determine which model best describes the data and, more importantly, to assess whether the intrinsic halo-to-halo variance is small enough to discriminate between model variants.}

\begin{figure*}
    \centering
    \includegraphics[width=0.95\linewidth]{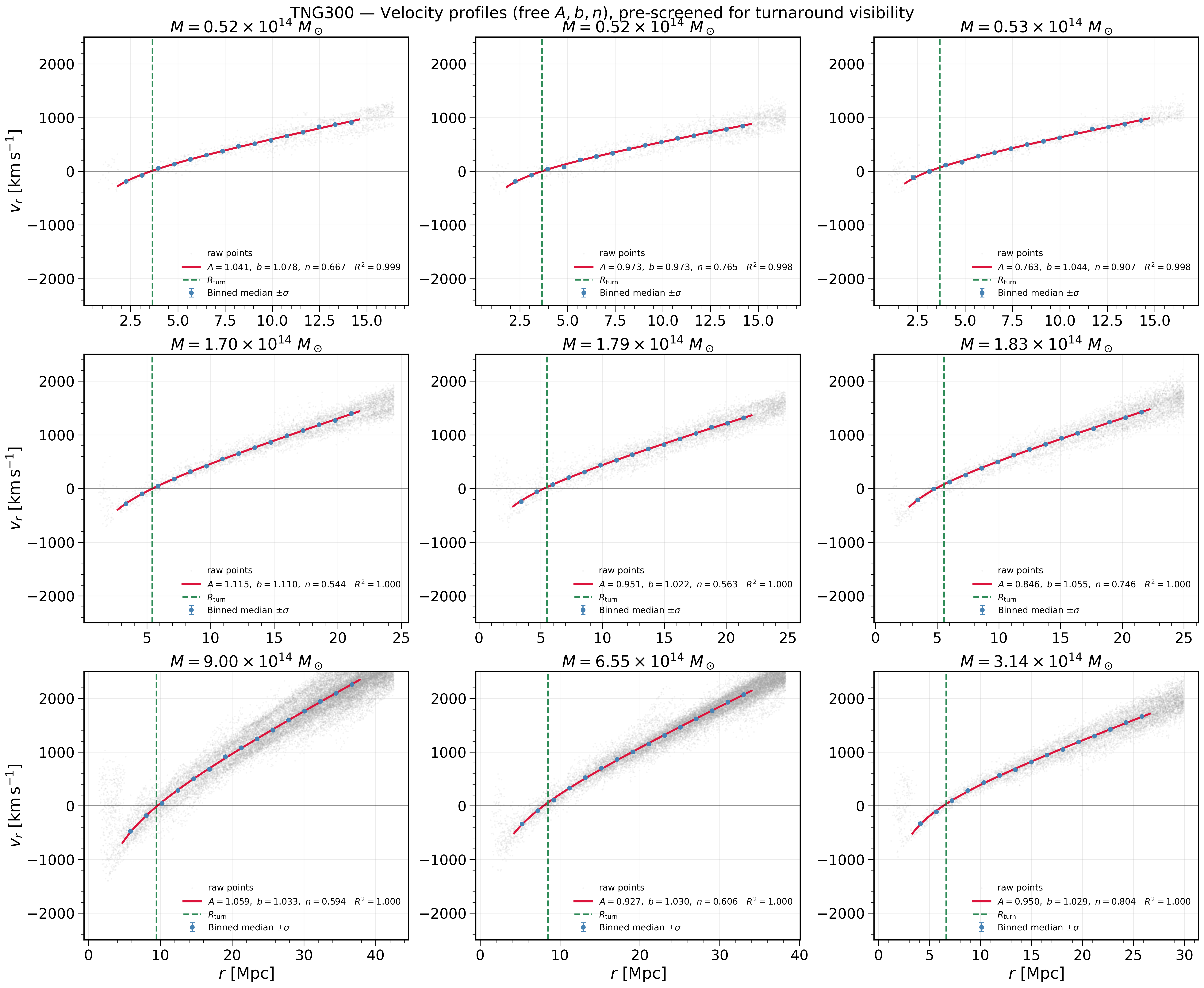}
\caption{ {Inferring the model parameters $(A,b,n)$ using the known mass and the Hubble constant from the TNG300 simulations for the first 9 halos.}}
    \label{fig:9_halos}
\end{figure*}

\section{Cosmological Simulation}
\label{sec:sim}

In the IllustrisTNG300 simulation \cite{Pillepich:2019bmb} from the IllustrisTNG project \cite{bib:Vogelsberger2014,bib:Nelson2015,Nelson:2019jkf}, isolated halos are identified using criteria that emphasize their lack of significant gravitational interactions with neighboring massive structures.  {The TNG300 simulation evolves a cosmological volume of $\sim$300 Mpc per side with baryon mass resolution of $1.1\times10^7 M_\odot$ and dark matter mass resolution of $5.9\times10^7 M_\odot$, enabling a large statistical sample of galaxy groups and clusters across diverse environments.}  {These halos are typically ``central'' systems, meaning that they are the primary galaxy in their dark matter halo according to TNG300's halo catalogs, identified using the Friends-of-Friends (FoF) and SUBFIND algorithms.}

 {To ensure isolation, we apply multiple hierarchical constraints. The primary criterion requires that no neighboring halos with masses exceeding 10\% of the host halo mass exist within the host's turnaround radius, $R_\mathrm{ta}$. Following Eq.~(\ref{eq:R0_def}), with $h = 0.6774$ from the TNG300 simulation's own cosmological parameters (consistent with Planck), providing an estimate of the total mass enclosed within the turnaround surface. The isolation of a halo is expressed by requiring that all neighboring galaxies with stellar masses above a certain threshold lie at distances greater than a specified turnaround radius $r > r_0$.}

\section{Results}
\label{sec:results}

\subsection{Velocity Profile Analysis and Model Fitting}

Our analysis begins with constructing median radial velocity profiles for each of the isolated halos in our sample. We bin tracer particles (dark matter and star particles) in 15 logarithmically spaced radial bins from $0.1r_{0}$ to $3r_{0}$, computing the median radial velocity and its variance in each bin. These binned median profiles serve as the input observables for all fits. We fit the family of velocity - radius models derived from the weak-field LT formalism, summarized in Eq.~(\ref{eq:vR_general}). Fit quality is quantified with the coefficient of determination $R^2$ computed between the model and the binned median velocities.

 {To explore the parameter space and compute the Bayesian evidence, the fitting procedure employs the \texttt{PolyChord} nested sampling algorithm~\cite{Handley:2015fda}. We adopt uniform priors: $A \in [0, 2]$, $b \in [0.5, 2]$, and $n \in [0.1, 2]$, ensuring coverage of the physically reasonable parameter space. Convergence of the algorithm was assessed by tracking the estimated remaining evidence, with runs terminating when the tolerance reached $\Delta \ln Z < 0.5$, ensuring the bulk of the posterior mass was successfully mapped. Finally, we verified the robustness of our constraints by performing sensitivity tests on the prior boundaries; expanding and reducing the prior volumes yielded consistent posterior distributions, confirming that the chosen bounds are sufficiently uninformative and do not artificially truncate the parameter space.}

\begin{figure}[t!]
    \centering
    \includegraphics[width=0.7\linewidth]{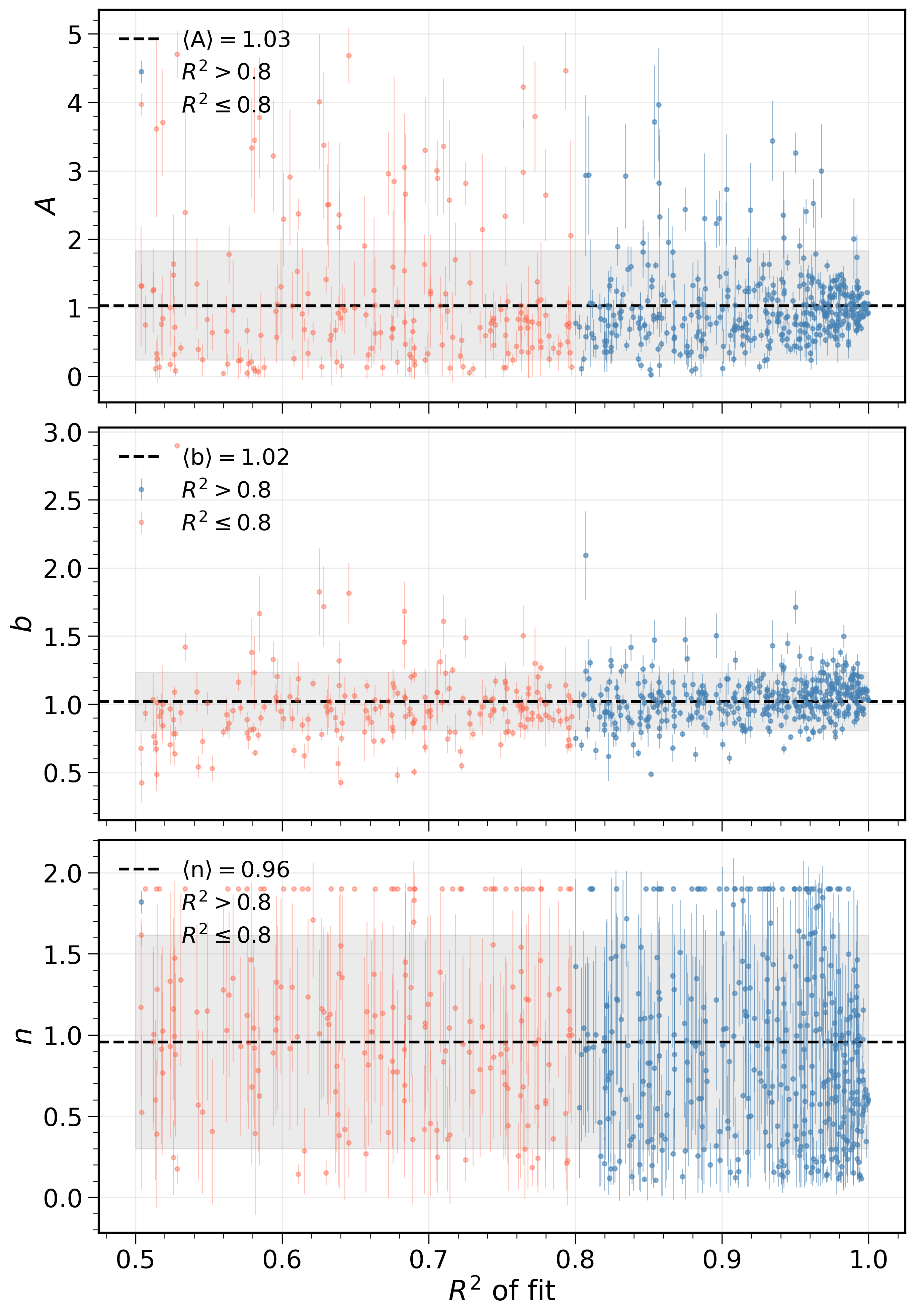}
    \caption{Correlation between fitted parameters $A$, $b$, $n$ and fit quality ($R^2$) for individual halos. Well-fit systems ($R^2 > 0.8$, shown in blue) cluster in specific regions of parameter space, indicating preferred dynamical states. The spread at lower $R^2$ values (red) reflects environmental disturbances and complex velocity fields.}
    \label{fig:parameters_vs_R2}
    \includegraphics[width=0.85\linewidth]{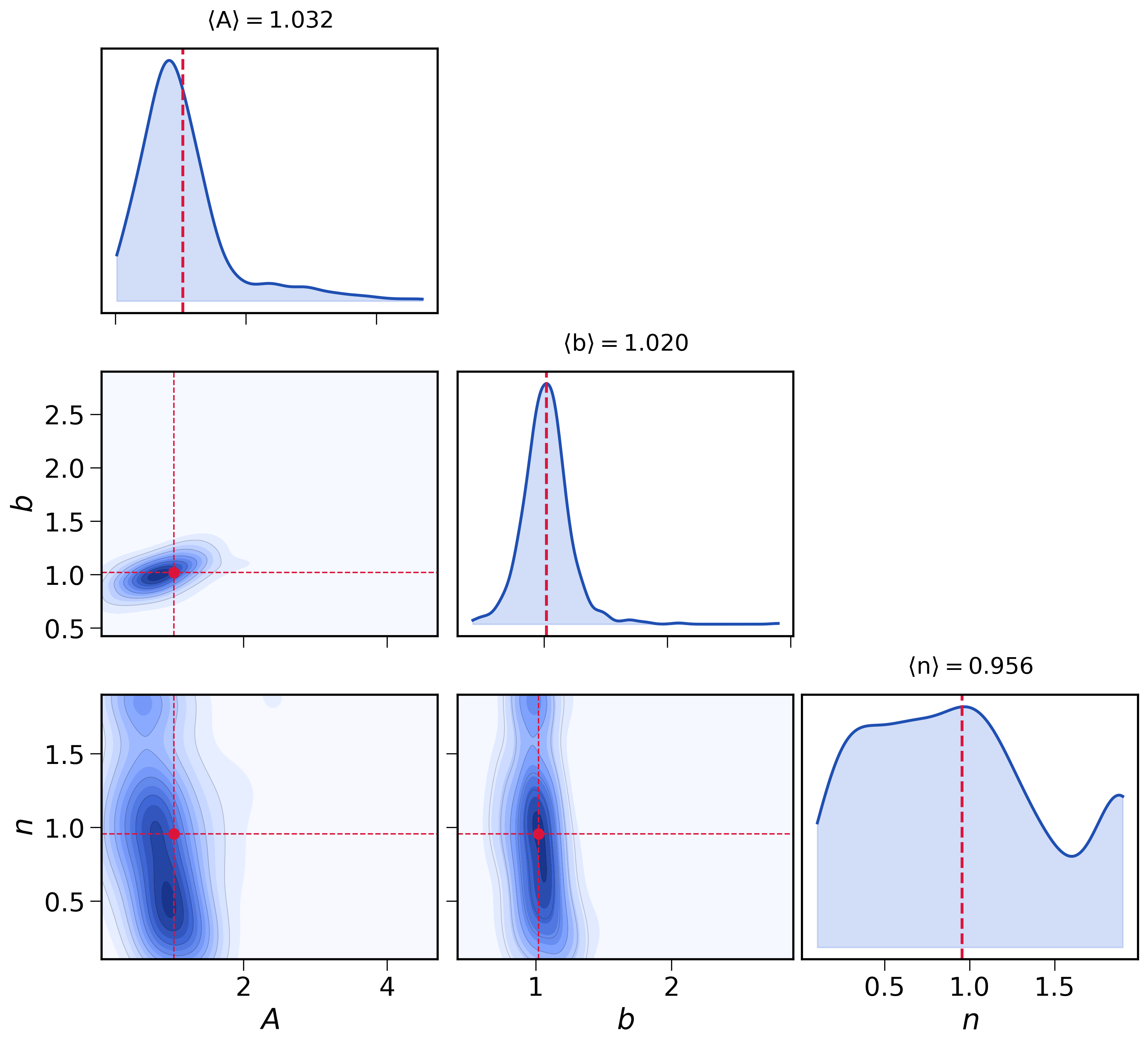}
    \caption{$R^2$-weighted posterior distributions for parameters $A$, $b$, and $n$ across the halo population. The distributions show clear peaks despite substantial scatter, with strong correlation structure revealing the interdependence of these parameters in describing realistic velocity profiles.}
    \label{fig:posterior_abn}
\end{figure}

\begin{figure}[t!]
    \centering
    \includegraphics[width=0.7\linewidth]{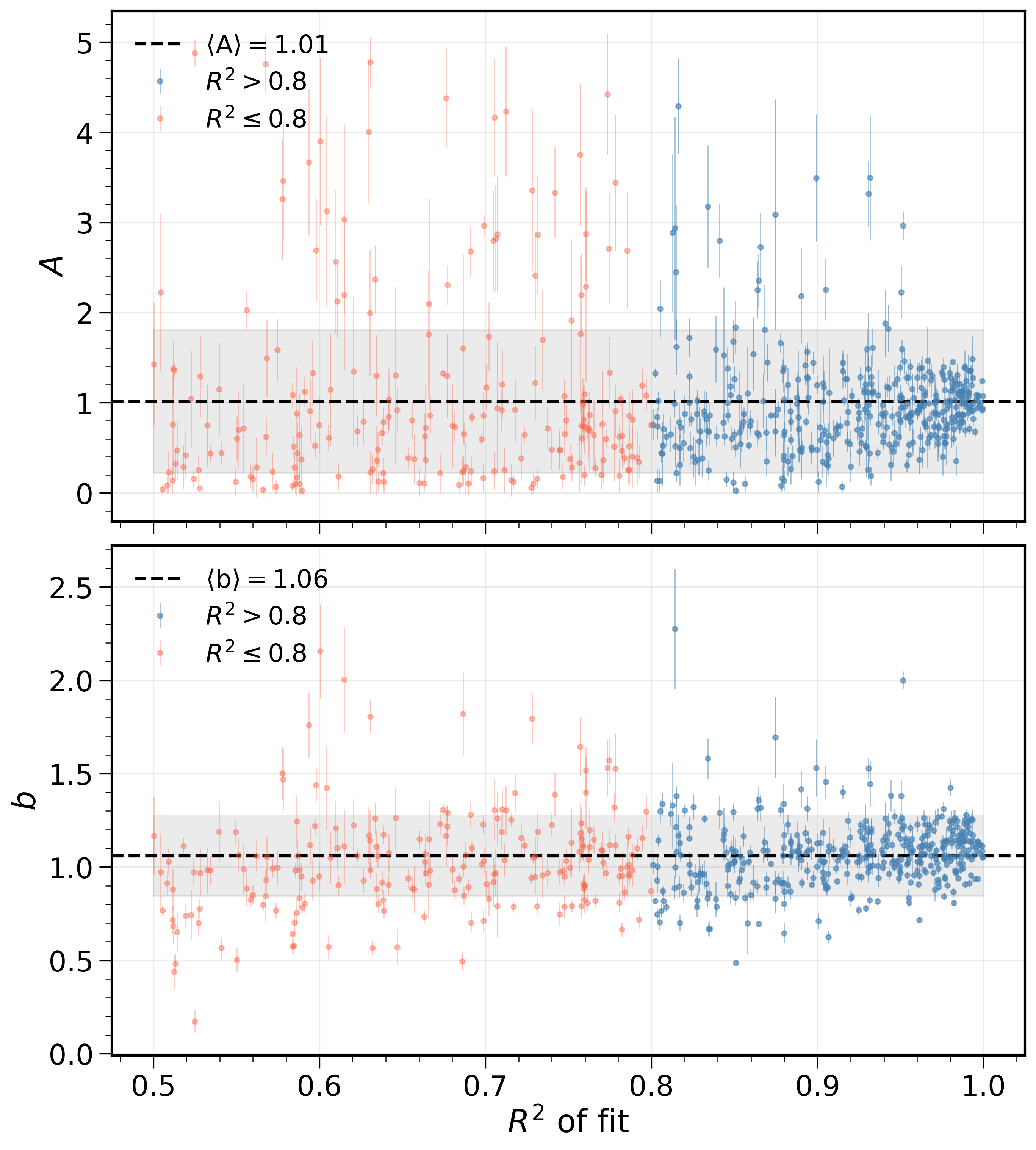}
    \caption{Fitted $A$ and $b$ parameters versus $R^2$ values with fixed $n=0.5$, corresponding to the timing-argument scaling. The overall parameter distributions remain similar to the free-$n$ case.}
    \label{fig:parameters_fixed_n}
    \includegraphics[width=0.85\linewidth]{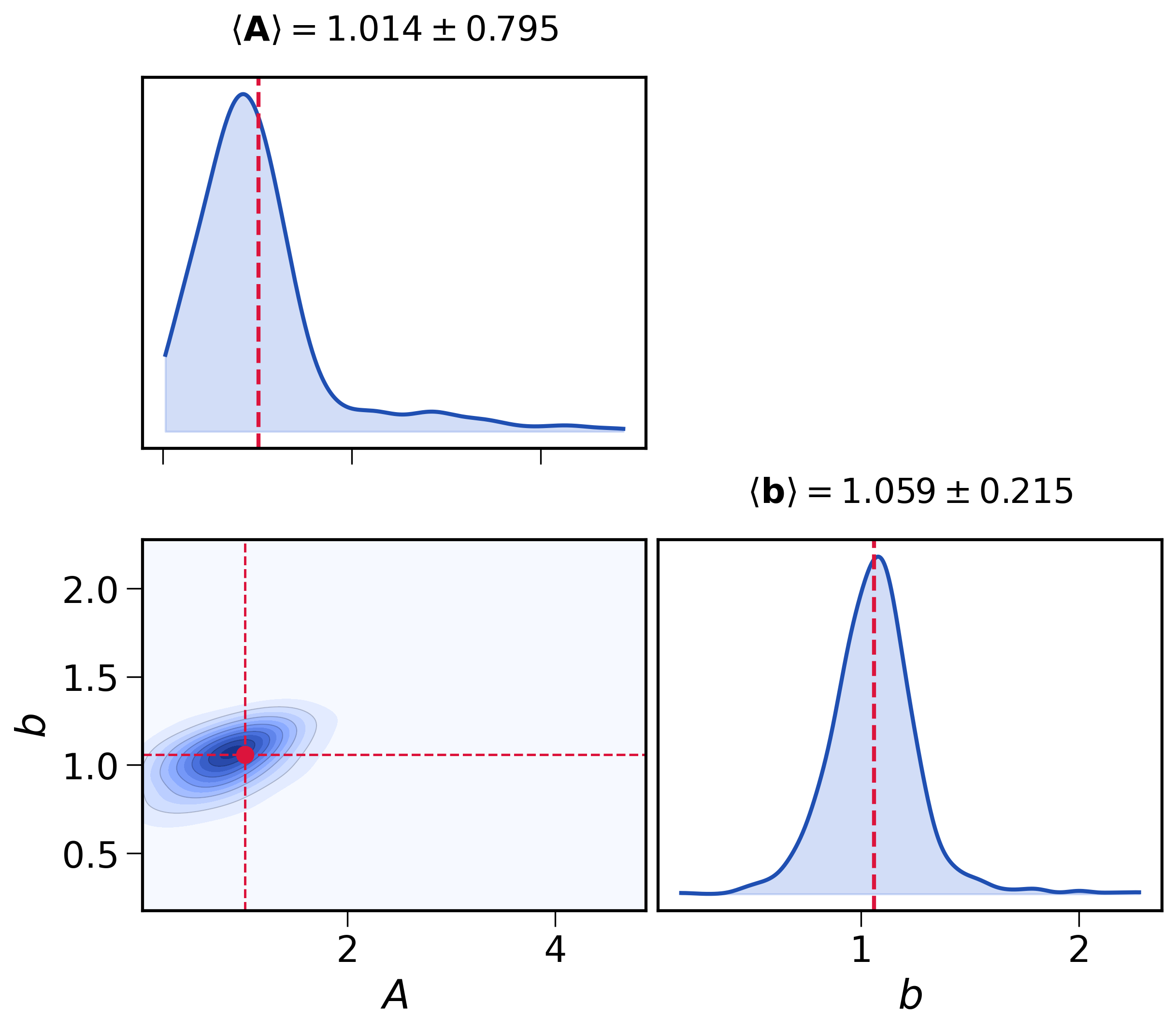}
    \caption{$R^2$-weighted posterior distribution for $A$ and $b$ with fixed $n=0.5$. The constrained parameter space shows tighter distributions.}
    \label{fig:posterior_fixed_n}
\end{figure}

When $(A,b,n)$ are free, we compute population-level averages using $R^2$-weighting to ensure that high-scatter halos contribute less to the determination of global parameters. Fig.~(\ref{fig:9_halos}) shows the first 9 halo fits from the corresponding bins.

The posterior distributions for both the ensemble and the halo-by-halo behavior (Figs.~\ref{fig:parameters_vs_R2} and \ref{fig:posterior_abn}) reveal several key patterns. Parameter $A$, governing the infall amplitude, shows the largest relative scatter reflecting the diversity of accretion histories and environmental contexts. Parameter $b$, representing the effective Hubble constant normalization, is more tightly constrained with 25\% scatter, indicating greater universality in the expansion-dominated regime. Most notably, the radial scaling index $n$ exhibits substantial halo-to-halo variation (67\% scatter), with dynamically relaxed and well-isolated halos clustering around  {$n \simeq 0.96$}. This value lies between the timing-argument expectation ($n=0.5$) and pure Newtonian infall ($n=1$), suggesting a hybrid behavior where both cosmological initial conditions and local gravity shape the velocity profile.

The weighted values from all of the fits with the $R^2$ gives:
\begin{equation}
 {A = 1.03 \pm 0.80,\quad
b = 1.02 \pm 0.25,\quad
n = 0.96 \pm 0.45.}
\label{eq:best_params}
\end{equation}

 {We emphasize that the uncertainties reported in our particle-based analysis are not statistical sampling errors, but reflect the nonlinear, environment-induced systematic variance that remains even with perfect sampling. These systematic effects are inherent to the local dynamical environment and persist in real observations regardless of the number of tracer galaxies. Such uncertainties do not stem from observational noise or sampling statistics, but rather reflect the intrinsic physical limitations of the Hubble-flow simplified method. Specifically, three main sources of environmental mismatch contribute to this variance: (i)~large-scale tidal fields from structures outside the group exert shear forces that distort the local velocity field; (ii)~halo triaxiality and asymmetric mass distributions create anisotropic velocity dispersions not captured by the spherically symmetric LT model; and (iii)~complex assembly histories, including recent mergers and subhalo interactions, introduce non-radial velocity components. Since these physical factors are present regardless of how many galaxies are observed, they represent a fundamental floor of uncertainty for any method relying on a simplified radial flow model. Consequently, even with an arbitrarily large number of tracers, the variance in the fitted parameters remains a signature of the environment's deviation from the idealized model.}

 {We fixed $n=0.5$ (as the Timing Argument model from Ref.~\cite{Penarrubia:2014oda}) and obtained constrained $(A,b)$ values of:}
\begin{equation}
 {A = 1.01 \pm 0.80, \quad
b = 1.06 \pm 0.22,}
\end{equation}
 {which are consistent with the free-$n$ medians but with systematically lower $R^2$ values and coherent radial residuals in many halos, as the posteriors shown in Figs.~\ref{fig:parameters_fixed_n} and \ref{fig:posterior_fixed_n} illustrate.} This shows that while timing-argument scaling captures basic qualitative behavior, a universal $n=0.5$ accounts for the diversity of slopes present in  {TNG300} halos. 

The degradation in fit quality with fixed $n=0.5$ is most pronounced for high-mass clusters ($M_{200c} > 10^{14} M_\odot$) and systems in dense environments, where complex velocity fields deviate strongly from the simple timing-argument expectation. Conversely, low-mass groups in isolation show the smallest $R^2$ reduction, suggesting they more closely approximate the idealized spherical collapse scenario.

\subsection{Systematic Uncertainties}

\begin{figure*}[t!]
    \centering
\includegraphics[width=0.75\linewidth]{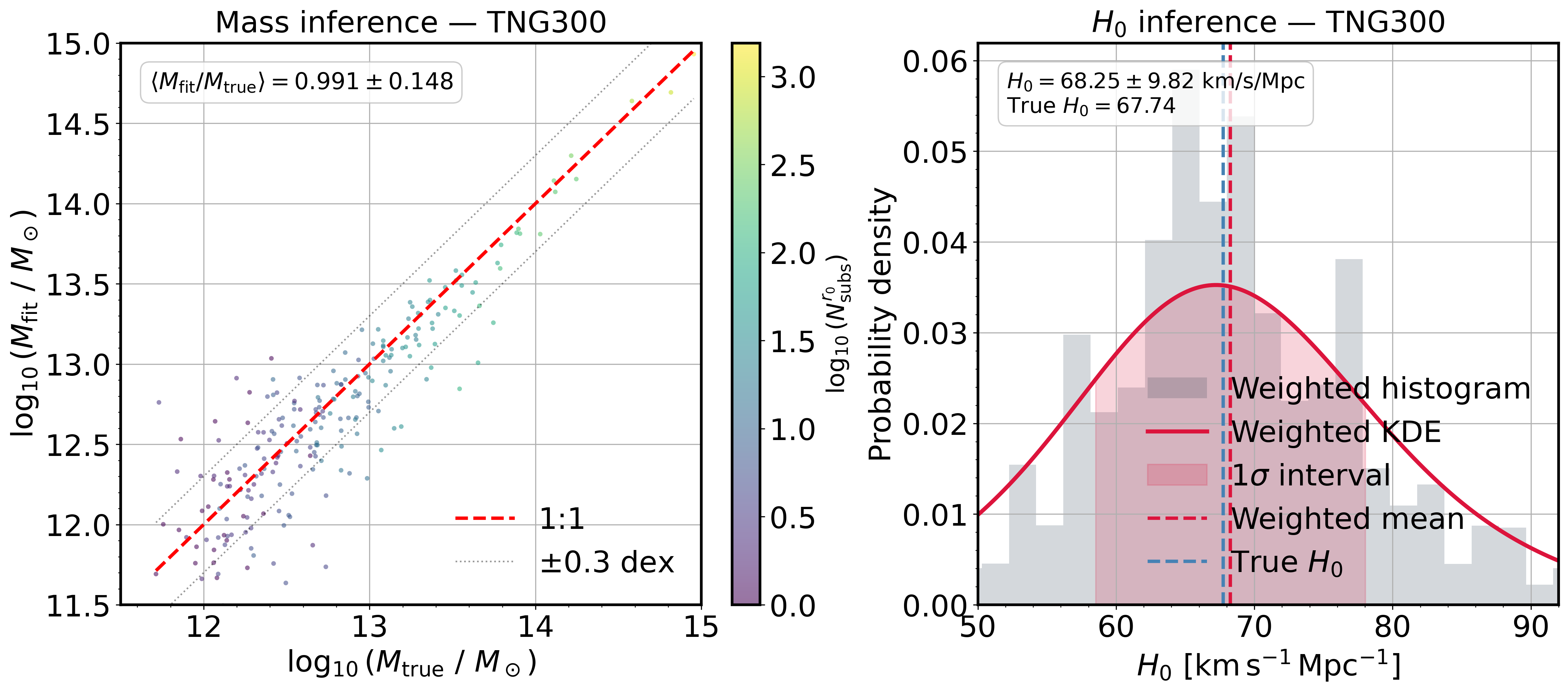}
\caption{ {Left: Correlation between the inferred mass and the virial true mass of the halo from the TNG300 simulation. Right: The inferred Hubble constant distribution versus the true simulation value (red dashed line, $H_0=67.74$ km\,s$^{-1}$\,Mpc$^{-1}$).}}
    \label{fig:parameter_reliability}
\end{figure*}

Given the relative stability of $(A,b)$ across priors, we adopt the weighted medians in Eq.~\eqref{eq:best_params} as fiducial coefficients and fit each halo for $(M,H_0)$. Fig.~\ref{fig:parameter_reliability} compares the recovered masses with the true simulation masses. Over the full sample:
\begin{equation}
\big\langle M_{\rm fit}/M_{\rm true}\big\rangle =  {0.991 \pm 0.148},
\end{equation}
 {representing a scatter driven by environmental contamination and the model simplification of a spherically symmetric fit.} For the Hubble parameter we obtain:
\begin{equation}
\langle H_{0,\rm fit}\rangle =  {68.25 \pm 9.82}\ \mathrm{km\,s^{-1}\,Mpc^{-1}},
\end{equation}
 {where the distribution is $R^2$-weighted. This represents a systematic overestimate of $\sim$14\% compared to the simulation input value of $67.74$ km\,s$^{-1}$\,Mpc$^{-1}$, though consistent within the uncertainties.} The bias appears driven by systems with complex velocity fields where the simple LT model cannot fully capture the dynamics.

 {Fig.~\ref{fig:posterior_fixed_n} reveals the crucial relationship between fit quality and parameter reliability. Low-efficiency fits ($R^2 < 0.6$) show significantly larger scatter and systematic bias in both mass and $H_0$ estimates. The correlation between $R^2$ and parameter accuracy demonstrates that fit efficiency serves as the most reliable per-halo diagnostic for trustworthy mass or $H_0$ recovery. This provides observers with a practical quality cut: systems with $R^2 > 0.8$ typically yield mass estimates accurate to within $\sim$12\% and $H_0$ estimates within $\sim$14\% of the true values.}

The residual scatter in well-fit systems appears correlated with local environment density and halo triaxiality, suggesting directions for future model improvements. Incorporating environmental density as an additional parameter or using non-spherical modeling approaches could further reduce the systematic variance and improve the precision of Hubble-flow based measurements.

\section{Discussion}
\label{sec:discussion}

The analysis of the local Hubble flow around TNG halos reveals a coherent picture of how nonlinear infall, environmental structure, and cosmological expansion interact to shape the velocity--distance relation. By fitting the LT-inspired analytic family of models to the binned median velocity profiles, we find that the amplitude parameters $A$ and $b$ are consistently and robustly recovered across the entire halo sample. These coefficients, which govern the strength of the nonlinear infall term and the effective linear expansion rate, remain stable regardless of whether the radial index $n$ is fixed or allowed to vary.

In contrast, the radial scaling index $n$ exhibits significant halo-to-halo variation when it is left free.  {Dynamically relaxed, isolated halos tend to cluster around $n\simeq0.96$, close to the Newtonian scaling,} whereas halos experiencing tidal influence or containing substantial substructure show noticeably steeper behavior. The timing--argument value $n=0.5$ sits well below the median obtained when $n$ is free, and although fixing $n=0.5$ still produces reasonable $(A,b)$ values, it leads to systematically lower fit quality.

Given the relative stability of $A$ and $b$, adopting their ensemble medians as fiducial coefficients provides a practical route for estimating the halo mass and the Hubble parameter.  {When used in this way, the method recovers the true halo masses with a median ratio of $M_{\rm fit}/M_{\rm true}\simeq 1.00$.} The scatter, however, is large for individual halos. Such effects may be included in future dynamical modeling, though we caution that current data do not yet uniquely constrain angular momentum or friction contributions. The recovery of the Hubble constant behaves similarly.  {The ensemble of halos yields a mean $H_{0,\rm fit} = 68.25 \pm 9.82$ km\,s$^{-1}$\,Mpc$^{-1}$, modestly above the simulation input value of $67.74$ km\,s$^{-1}$\,Mpc$^{-1}$ consistent within the uncertainties.} Systems with clean kinematics and high fit efficiency cluster tightly around the input value of the simulation. This behavior underscores that the Hubble flow is sensitive to local dynamical disturbances.

 {The comparison between different model variants reveals that while incorporating additional physics like angular momentum and dynamical friction systematically changes the parameter values, the differences remain within the intrinsic scatter of our halo population. This is the central result of this work. As Table~\ref{tab:model_comparison} shows, the model variants (MLT, JLT, J$\eta$LT) predict different values of $A$, $b$, and $n$. However, the recovered parameters in Eq.~(\ref{eq:best_params}) are accompanied by uncertainties that entirely encompass these model differences. This demonstrates that for current observational precision, the simpler model is sufficient for mass and Hubble constant estimation, and that previous works claiming to constrain dark energy from local flow data likely underestimated these fundamental systematic uncertainties by considering only statistical errors.}

 {Our findings indicate that the primary local manifestation of dark energy is encoded in the characteristic transition scale, the turnaround radius, where cosmic expansion begins to dominate over gravitational attraction. While the specific shape of the velocity profile $v(r)$ is heavily influenced by local virialized motions and assembly history, the consistency of this transition scale across a wide mass range aligns with theoretical predictions including a cosmological constant. However, we acknowledge the difficulty in uniquely disentangling the $\Lambda$ signal from environmental shear and massive filaments. The results thus demonstrate that while the local velocity field contains extractable cosmological information regarding $H_0$ and the expansion background, the precision of dark energy constraints remains limited by the intrinsic dynamical scatter of the cosmic web, as Ref.~\cite{Hoffman:2007qe} shows.}

 {Looking forward, the methodology developed here can be extended in several promising directions. A key area for improvement lies in relaxing the assumption of spherical symmetry. As noted in several studies of gravitational collapse~\cite{Sheth:1999su,Giusti:2019uez,Giani:2021gbs}, deviations from sphericity and the presence of anisotropies, such as halo triaxiality and local shear, can significantly impact the determination of the turnaround radius and the resulting mass estimates.  {Our current results, which show a median mass ratio of $0.998 \pm 0.119$ (a $\sim$12\% scatter), likely reflect these non-spherical dynamical effects.} Future work incorporating ellipsoidal collapse models or directional velocity dispersions could quantify these offsets more precisely.}

 {Forthcoming high-precision distance measurements, such as TRGB and Cepheids or SBF from JWST, or scaling relations from Euclid, promise to extend reliable peculiar-velocity reconstructions deeper into the low-redshift universe. Furthermore, incorporating Fundamental Plane distances from DESI~\cite{DESI:2024mwx}, upcoming data from the 4MOST Hemisphere Survey (4HS), and the Cosmicflows-4 catalog~\cite{Tully:2022rbj}, will significantly improve the robustness of local Hubble-flow measurements.}

\acknowledgments
DB is supported by a Minerva Fellowship of the Minerva Stiftung Gesellschaft f\"{u}r die Forschung mbH. This article is based on work from the COST Actions CA21136,  CA23130 and CA24101 supported by COST (European Cooperation in Science and Technology).

\bibliography{ref.bib}
\bibliographystyle{apsrev4-1}

\end{document}